\documentclass[10pt,conference,british]{IEEEtran} 

\usepackage{blocked}
\usepackage{color}
\usepackage{multirow}
\usepackage{subcaption}
\usepackage{soul}
\usepackage {amsmath} 
\usepackage{stfloats} 
\usepackage{makecell} 
\usepackage{graphicx}
\usepackage[bookmarks=false]{hyperref}
\hypersetup{nolinks=true}
\usepackage{booktabs}
\usepackage{comment}
\usepackage{amsmath, amssymb}
\usepackage{cite}

\begin{document}

\title{Stablecoins as Dry Powder: A Copula-Based Risk Analysis of Cryptocurrency Markets}

%
%

\author{
\IEEEauthorblockN{Elliot Jones}
\IEEEauthorblockA{\textit{Department of Computing} \\
\textit{Imperial College London}\\
United Kingdom\\}
\and
\IEEEauthorblockN{Toshiko Matsui}
\IEEEauthorblockA{\textit{Department of Computing} \\
\textit{Imperial College London}\\
United Kingdom\\}
\and
\IEEEauthorblockN{William Knottenbelt}
\IEEEauthorblockA{\textit{Department of Computing} \\
\textit{Imperial College London}\\
United Kingdom\\}
}
\IEEEoverridecommandlockouts
\maketitle              
%
\begin{abstract}

Stablecoins serve as the fundamental infrastructure for Decentralised Finance (DeFi), acting as the primary bridge between fiat currencies and the digital asset ecosystem. While peg stability is well-documented, the structural role stablecoins play in transmitting systemic risk to the broader market remains under-explored. This study uses copula-based approaches to quantify the transmission of volatility and activity from stablecoin to cryptocurrency markets. We demonstrate in-sample causality across daily, weekly, and monthly horizons. Furthermore, we show that incorporating stablecoin factors significantly reduces Mean Squared Error in cryptocurrency forecasting. Specifically, we link stablecoin volume and upside volatility to broader market volatility, indicating its role as dry powder. Finally, we establish economic value by demonstrating reduced risk in a cryptocurrency volatility targeting model when stablecoin factors are employed.

\end{abstract}

%

\section{Introduction}
Stablecoins are becoming an increasingly important entity in Decentralised Finance (DeFi) for both investors and policymakers. This is reflected by the introduction of new regulatory frameworks in various countries, such as the GENIUS Act in the United States~\cite{GENIUS2025} and the Stablecoin Ordinance in Hong Kong~\cite{HKStablecoin2025}. In fact, stablecoins have become increasingly popular over the last year~\cite{NBER-SC2025}, with their market cap reaching \$317 billion~\cite{coinmarketcapStables}, around a seventy-fold increase since early 2020. Stablecoins are taking on an increasingly important role in the digital asset ecosystem, providing a stable medium of exchange within the volatile crypto environment~\cite{Gorton2023taming}.

Much of this prominence stems from their peg mechanisms, which make stablecoins less volatile than traditional cryptocurrencies~\cite{Almeida2024Crypto, matsui2022bgo_dynamics}. Prior empirical research on stablecoins is scarce~\cite{Ante2023systematic,Dionysopoulos2024sytematic}, examining areas such as price and volume correlations with other assets, stability and volatility dynamics, and the influence of macroeconomic factors. However, there remains no consensus on the presence of systemic risk transmission from stablecoins to cryptocurrencies~\cite{Mahrous2025SC}.

Therefore, this study investigates the effect stablecoins have on cryptocurrency markets. We employed copula-based approaches for both In-Sample and Out-of-Sample (OOS) tests, allowing us to assess non-linear relationships across the whole joint distribution of the two markets. In particular, we use Copula Granger Causality (CGC)~\cite{copulaGranger} to test for in-sample causality and a novel GARCH-Copula-XGBoost Framework to forecast volatility in cryptocurrency markets. To capture the market-wide effects, we use Principal Component Analysis (PCA) to obtain market factors for both stablecoin and cryptocurrencies. We also split volatility into its upside and downside components so that we can see the effect of stablecoins on both sides of the market individually.

Using data from 1 January 2020 to 1 January 2025, covering the three largest stablecoins (“3Pool”) and the four largest cryptocurrencies, our analysis\footnote{All code, datasets and plots from this research are available online at: \href{https://github.com/EllbellCode/StablecoinAnalysis}{github.com/EllbellCode/StablecoinAnalysis}} reveals significant causal relationships, reductions in predictive error, and improvements in risk-related performance metrics when employing stablecoin data in our models. Namely, we show that accumulation (upside volatility) and activity (volume) in stablecoins lead cryptocurrencies, legitimising the idea of stablecoins acting as a "Dry Powder" for cryptocurrency markets. Finally, we briefly touch upon recent regulatory acts and how they are likely to shape the role of stablecoins.


\label{sec:Intro}



\section{Literature Review}

Past research has investigated the stablecoin price/volume correlation with other assets. Among the earliest work in this domain, stablecoins are shown to exhibit low or negative correlation with both cryptocurrencies and traditional assets, especially when the market is under stress~\cite{Mahrous2025SC}. This stability, or safe haven nature, is most noticeable during times of extreme volatility, such as the COVID-19 outbreak~\cite{BAUR2021safe,KLIBER2022safe}.

On the influence on other assets, stablecoins such as TrueUSD, USD Coin, and Tether are observed to exhibit a bidirectional, time‑varying causal relationship with crude oil prices~\cite{GHABRI2022oil}.
Regarding the interaction between stablecoins and cryptocurrencies, \cite{Grobys2022Jump} showed using Barndorff–Nielsen and Shephard (BNS) model~\cite{Barndorff2005BNM} that jumps in Tether prices predict a drop in bitcoin prices. \cite{PAENG2024VAR} and~\cite{GUBAREVA2023QQR} have also documented bidirectional causality and spillover effects between stablecoins, major cryptocurrencies, and traditional financial assets, by using a Quantile Vector Autoregression model to perform Granger Causality.
Conversely, \cite{chen2022vol} identified minimal causality and spillover effects between stablecoins and cryptocurrencies, by employing a combination of GARCH and VAR models.

Methodology wise, GARCH models and their variants have been the predominant methods employed (e.g. \cite{WANG2020GARCH,FENG2024GARCH}). However, copula models better capture
non-linear relationships than GARCH models. \cite{BELGUITH2024Copula} demonstrated by employing time-varying Student's copula that the degree of dependence between gold-backed cryptocurrencies and NFT and DeFi tokens change over time and that gold-backed cryptocurrencies serve as hedging assets. \cite{chen2022vol} showed by introducing several range-based volatility estimators to the BEKK- GARCH and Copula-DCC-GARCH that bitcoin and stablecoin market volatility are interconnected. Machine learning-based approaches have seen frequent use in cryptocurrency time series analysis, including regressions~\cite{regressionComparative}, trees~\cite{xgBoostCryptoPrice} and neural networks~\cite{lstmCryptoPrice}. However, most literature focuses on predicting returns and mostly do not use hybrid framework approaches with the exception of Autoregressive Moving Average (ARMA) models~\cite{machinelearningARMA}.
\label{sec:Lit}
\section{Preliminaries} 
\label{sec:preliminaries}

\subsection{Volatility}
\label{subsec:vol}

The Rogers--Satchell volatility estimate~\cite{Rogers1991}, a range-based estimator consisting of an upside and downside component, is defined as follows:
\begin{align}
    \sigma_t^2 = \ln \left( \frac{H_t}{C_t} \right) \ln \left( \frac{H_t}{O_t} \right) + \ln \left( \frac{L_t}{C_t} \right) \ln \left( \frac{L_t}{O_t} \right)
\end{align}

where $O_t, H_t, L_t, C_t$ represent the open, high, low, and close prices on day $t$. To obtain the upside and downside volatility components, we take the square root of each term:
\begin{align}
    \sigma_{\mathit{up,t}} &= \sqrt{\ln \left( \frac{H_t}{C_t} \right) \ln \left( \frac{H_t}{O_t} \right)} \\
    \sigma_{\mathit{down,t}} &= \sqrt{\ln \left( \frac{L_t}{C_t} \right) \ln \left( \frac{L_t}{O_t} \right)}
\end{align}

We choose the Rogers--Satchell estimate as it enables us to split volatility into its upside and downside components to see how stablecoins influence both sides of crypto price movements. Moreover, the model accounts for a trend in the data, a property almost always exhibited by crypto markets~\cite{RangeBasedVolCrypto}. Another popular range-based estimator, often deemed superior to Rogers-Satchell, is the Yang--Zhang volatility~\cite{yangzhangVol}. However, its main advantage over Rogers--Satchell is its ability to account for volatility between trading sessions, an issue mitigated by the fact that crypto markets trade $24/7$. Outside of range-based estimators, another common measure of volatility is realised volatility (RV)~\cite{Andersen1998RV}. However, range-based volatility estimators have been shown to be more accurate and efficient than RV, which are usually return-based~\cite{Parkinson1980,garman1980estimation}. Furthermore, range-based volatility estimators are efficient and robust to microstructure noise such as bid-ask bounce~\cite{Alizadeh2002range}.

\subsection{GARCH}
\label{subsec:garch}

Generalised Auto-Regressive Conditional Heteroskedasticity (GARCH) is a statistical model designed to estimate returns and volatility in financial markets, with a specific focus on addressing volatility clustering~\cite{garch}. We use a variation where both the mean and variance are autoregressive, depending on their own past values. For simplicity, we set the order of autoregression to one, meaning today's values depend only on yesterday's values. Furthermore, we extend the GARCH model with E-GARCH~\cite{egarch} to account for the leverage effect~\cite{leverageEffect}, the asymmetric volatility seen in cryptocurrency markets (and almost all other financial markets) as a result of downside volatility being larger than upside volatility~\cite{leverageCrypto}. The equations for our E-GARCH model are given as follows:
\begin{align}
    r_t &= \mu + \phi r_{t-1} + \epsilon_t \\
    \epsilon_t &= \sigma_t z_t, \quad z_t \sim \textit{SkewT}(\mu_{\mathit{dist}}, \sigma_{\mathit{dist}}, \nu, \xi) \\
    \sigma_t &= \sqrt{\exp \left( \omega + \beta \ln(\sigma_{t-1}^2) + \gamma z_{t-1} + \alpha \left( \left| z_{t-1} \right| - E[|z_t|] \right) \right)}
\end{align}
where $r_t$, $\epsilon_t$, $z_t$ are the returns, error and shock of day $t$ respectively. Furthermore, we have the trend coefficient $\mu$, autoregressive return coefficient $\phi$, Skewed Students-t distribution $\textit{SkewT}$ (with mean $\mu_{\mathit{dist}}$, standard deviation $\sigma_{\mathit{dist}}$, $\nu$ degrees of freedom, and $\xi$ skew), constant volatility $\omega$, magnitude $\alpha$, persistence $\beta$ and leverage $\gamma$.

Standard GARCH models usually have a normal distribution to model shocks. Instead, we use a Skewed Student's t distribution as this allows us to account for the heavy tails and asymmetry present in crypto markets~\cite{heavyTailsCrypto}.

\subsection{Copulas}
\label{subsec:copula}

A Copula links univariate marginal distributions to form a multivariate joint distribution function, isolating the dependence structure between variables from their individual behaviours~\cite{copulaModelling}. They are grounded in Sklar's Theorem~\cite{sklarstheorem}, which states that any multivariate joint distribution function $F$ can be expressed in terms of univariate marginals $F_i$ and a copula function $C$ that describes the dependence between variables $x_i$ in a $d$-dimensional space:
\begin{equation} F(x_1, \dots, x_d) = C(F_1(x_1), \dots, F_d(x_d)) \end{equation}

Crucially, all input variables to $C$ must be distributed equally to prevent individual marginal distributions from influencing the data. To ensure this, we use the Probability Integral Transform (PIT) to map all variables to a uniform distribution between 0 and 1. This means the copula only works on the relative ranks of our variables, isolating interdependence. 


Copulas enable us to model non-linear relationships in all parts of the distribution without influence from the underlying marginal distributions. This captures how assets move together both on average and in the tails of their distributions.

\subsection{XGBoost}
\label{subsec:xgboost}

XGBoost is an ensemble machine learning model that uses sequential decision trees where each subsequent tree is trained on the prediction errors of the previous tree~\cite{xgboost}. The forks in a tree are decided based on the calculation of Gain, which measures the reduction in loss provided by a specific feature.

For this study, XGBoost is the superior choice over common neural network architectures such as Long Short Term Memory (LSTM) and Transformer models. Daily data is limited in sample size, making our dataset unsuitable for a neural network that requires large datasets~\cite{treeModelsBetterThanDLforTabular,deepLearningBadTabular}. Furthermore, neural networks are black-box in nature, making it difficult to interpret the role of our variables. XGBoost, by contrast, is highly effective on smaller datasets, using of Gain making it possible to quantify the importance of our variables ~\cite{treeModelExplainability}.

\subsection{Diebold--Mariano}
\label{subsec:Diebold}

When comparing performance between benchmark and challenger models, it is insufficient to show that the challenger model has reduced error as this result could just be an anomaly. The Diebold--Mariano (\textit{DM}) test offers a way to determine if the performance difference is statistically significant~\cite{dieboldmariano}.

Let $d_t$ be the loss differential between two models at time $t$. The \textit{DM} test evaluates the null hypothesis that the loss differential is zero, with the statistic is calculated as $DM = \frac{\Bar{d}}{\hat{\sigma}_{\Bar{d}}}$, where $\Bar{d}$ is the sample mean of the loss differential and ${\hat{\sigma}_{\Bar{d}}}$ is its standard error. We apply two critical adjustments to account for the specific characteristics of our data and model structure.

First, we apply the Harvey--Leybourne--Newbold (HLN) correction, adjusting the test statistic based on the sample size $T$ and forecast horizon~\cite{DMTestHLN}. Tests on small to moderate sample sizes often reject the null hypothesis too easily, so the HLN correction forces our test to be more strict and reduce spurious results. We set the horizon to $1$ to forecast one day ahead and the correction simplifies to:
\begin{equation}
DM_{HLN} = DM \times \sqrt{\frac{T - 1}{T}}
\end{equation}

Next, we employ the Clark--West (CW) adjustment~\cite{DMTestClarkeWest} to account for the nested nature of our models. In the context of our research, the benchmark model is trained on cryptocurrency data only and our challenger model is trained on both cryptocurrency and stablecoin data. For this reason, the challenger model nests the benchmark model. Under the null hypothesis, the additional stablecoin features are irrelevant, meaning the larger model is expected to increase error as a result of the increased number of parameters. The CW adjustment corrects for this bias by modifying the loss differential:
\begin{equation} d_{CW, t} = e_{B,t}^2 - \left[ e_{C,t}^2 - (\hat{y}_{B,t} - \hat{y}_{C,t})^2 \right] \end{equation}
where $\hat{y}_{X,t}$ and $e_{X,t}$ are the predicted value and error from model $X$ at time $t$ respectively. 

\subsection{Sortino Ratio}

The Sortino Ratio is a way of measuring risk-adjusted returns~\cite{sortinoRatio}. The equation is as follows:
\begin{equation}
S = \frac{R_p - r_f}{\sigma_d}    
\end{equation}
where $R_p$ is the annualized portfolio return, $r_f$ is the risk-free rate (Minimum Acceptable Return) and $\sigma_d$ is the downside deviation, calculated using only the returns that fall below $r_f$. For this research we set the risk-free rate to 0, focusing on absolute loss~\cite{sortinoRatioZero}.

Unlike the Sharpe Ratio, which penalises both upside and downside volatility, the Sortino Ratio penalises only downside volatility as upside volatility is not considered a risk.
\section{Data}
\label{sec:data}

\subsection{Data Selection}

We use five years of daily price and volume data, spanning from 1st January 2020 to 1st January 2025. For stablecoins, we select DAI, USDC, and USDT, which represent the three largest stablecoins by market capitalisation and account for approximately 84\% of stablecoin market capitalisation~\cite{coinmarketcapStables}. These coins comprise the liquidity pool "3Pool" in decentralised finance (DeFi) protocols~\cite{curve3poolwebsite}. Although USDe currently commands a higher market capitalisation than DAI, it was excluded due to its 2024 release, offering insufficient data. For cryptocurrencies, we select BTC, ETH, BNB, and XRP, the four largest cryptocurrencies by market capitalisation and account for approximately 75\% of cryptocurrency market capitalisation excluding stablecoins~\cite{coinmarketcapTopCryptos}. All data was obtained from \textit{Investing.com}~\cite{investingdotcom}, containing the daily Open, High, Low, Close and Volume for each coin. All timestamping is denominated in Coordinated Universal Time (UTC), where the daily trading session opens at 00:00:00 and closes at 23:59:59.

\subsection{Metrics and Stationarity}

In time series analysis, it is essential that all time series are stationary, meaning their statistical properties remain constant over time~\cite{stationarityInTimeSeries}. We therefore transform our data into the first difference of the log volume $\Delta v_t$, and the first differences of the Rogers-Satchell upside and downside volatility components $\Delta \sigma_{U,t},\Delta \sigma_{D,t}$ defined in Section \ref{subsec:vol}. The calculations for these metrics are summarised below:
\begin{align}
\Delta v_t = \ln\left(\frac{V_t}{V_{t-1}}\right) \\
\Delta \sigma_{\mathit{up, t}} = \sigma_{\mathit{up, t}} - \sigma_{\mathit{up, t-1}}\\
\Delta \sigma_{\mathit{down, t}} = \sigma_{\mathit{down, t}} - \sigma_{\mathit{down, t-1}}
\end{align}

To confirm stationarity, we ran an Augmented Dickey--Fuller (ADF) test~\cite{ADF1979} on each time series. As shown in Table \ref{tab:adf_summary}, we obtain very low p-values for every series, showing stationarity.

\begin{table}[htbp]
    \centering
    \captionsetup{font=small}
    \caption{Summary of Augmented Dickey--Fuller (ADF) Test Results.}
    \resizebox{\columnwidth}{!}{
        \begin{tabular}{lcc}
            \toprule
            \textbf{Variable} & \textbf{ADF Statistic (Range)} & \textbf{Max $p$-value} \\
            \midrule
            $\Delta v_t$       & -11.79 to -10.33 & $< 0.0001$ \\
            $\Delta \sigma_{\mathit{up, t}}$ & -16.49 to -12.82 & $< 0.0001$ \\
            $\Delta \sigma_{\mathit{down, t}}$ & -14.58 to -13.22 & $< 0.0001$ \\
            \bottomrule
        \end{tabular}
    }
    \label{tab:adf_summary}
\end{table}

\subsection{Data Preprocessing}

Prior to modelling, we apply winsorisation, the process of bounding extreme outliers to a quantile~\cite{winsorisation}, to all time series at the $1\%$ and $99\%$ quantiles. The sample period encompasses multiple "black swan" events, extreme market shocks driven by macroeconomic factors rather than market dynamics. Notable examples include the Terra (LUNA) collapse~\cite{terraLunaColapse}, the FTX exchange failure~\cite{ftxCollapse}, and the Silicon Valley Bank crisis~\cite{svbCollapse}. Winsorisation mitigates the risk of these events skewing our test and models, ensuring they capture the true market dynamics rather than fitting on single events. Winsorisation is only applied to our training data whilst the testing data remains strictly unwinsorised to evaluate model robustness against these black swan events.

\subsection{Principal Component Analysis}

We employ PCA~\cite{PCA} to obtain cryptocurrency and stablecoin market factors. This enables us to capture the market-wide picture and reduce spurious results from testing individual coins and variables.

We apply PCA separately to each variable category. This dimensionality reduction transforms our individual asset time series into five market factors: 

\begin{enumerate} 
\item Stablecoin Volume $v_{\mathit{S}}$
\item Stablecoin Upside Volatility $\sigma_{\mathit{S, up}}$
\item Stablecoin Downside Volatility $\sigma_{\mathit{S, down}}$
\item Cryptocurrency Volume $v_{\mathit{C}}$
\item Cryptocurrency Upside Volatility $\sigma_{\mathit{C, up}}$
\item Cryptocurrency Downside Volatility $\sigma_{\mathit{C, down}}$
\end{enumerate}

The coin loadings of each factor's first principal component are given in Table \ref{tab:pca_loadings_combined}.

\begin{table}[htbp]
    \centering
    \caption{First Principal Component (PC1) Loadings.}
    \setlength{\tabcolsep}{1.5pt} 
    
    \resizebox{\columnwidth}{!}{%
        \begin{tabular}{l ccc c cccc}
            \toprule
            & \multicolumn{3}{c}{\textbf{Stablecoins}} & & \multicolumn{4}{c}{\textbf{Cryptocurrencies}} \\
            \cmidrule(r){2-4} \cmidrule(l){6-9}
            \textbf{Factor} & \textbf{DAI} & \textbf{USDC} & \textbf{USDT} & & \textbf{BNB} & \textbf{BTC} & \textbf{ETH} & \textbf{XRP} \\
            \midrule
            $v$ & 0.5670 & 0.5911 & 0.5737 & & 0.5009 & 0.5377 & 0.5441 & 0.4049 \\
            $\sigma_{\mathit{up}}$ & 0.5473 & 0.5994 & 0.5842 & & 0.4968 & 0.5290 & 0.5533 & 0.4089 \\
            $\sigma_{\mathit{down}}$ & 0.6121 & 0.5524 & 0.5658 & & 0.4912 & 0.5099 & 0.5203 & 0.4774 \\
            \bottomrule
        \end{tabular}%
    }
    \label{tab:pca_loadings_combined}
\end{table}

To select which principal components to keep, we use Horn's Parallel Analysis~\cite{hornAnalysis}, a statistical method that compares the eigenvalues $\lambda_{corr}$ of the correlation matrix against the eigenvalues derived from a correlation matrix of randomly-generated data. A principal component is retained only if its eigenvalue exceeds the corresponding critical eigenvalue $\lambda_{crit}$ generated from the random correlation matrix at the $95\%$ level.

The results are summarised in Table \ref{tab:pca_horn_results}. Our results indicate that for all factors, only the first principal component (PC1) explains significantly more variance than random noise. Consequently, we retain only the first component for each factor. For example, {Fig.  \ref{fig:scree_crypto_upside} shows Horn's analysis on the principal components of cryptocurrency upside volatility.

\begin{figure}[htbp]
    \centering
    \includegraphics[width=\columnwidth]{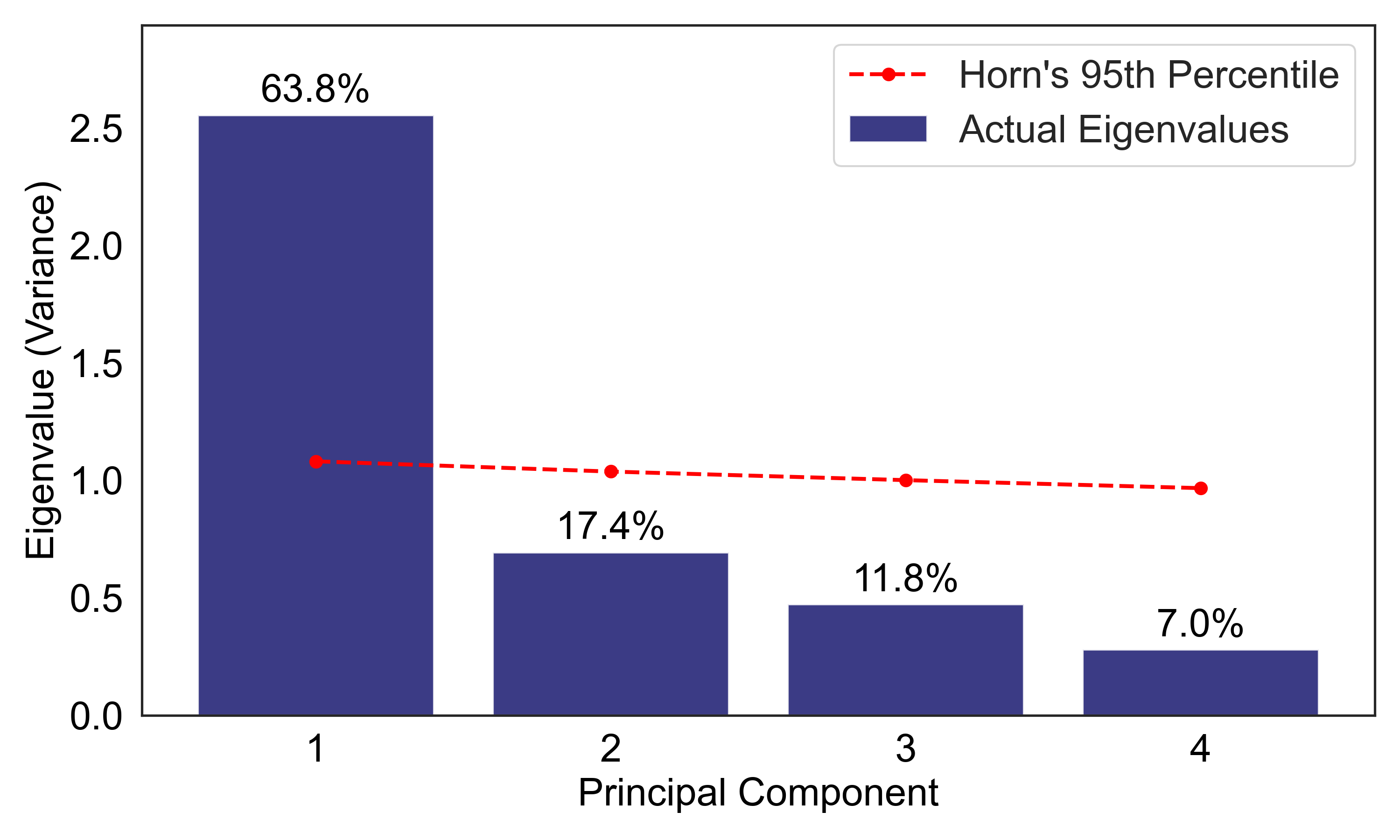} 
    \caption{Scree Plot for Cryptocurrency Upside Volatility Factor.}
    \label{fig:scree_crypto_upside}
\end{figure}

As shown in Table \ref{tab:pca_loadings_combined}, every stablecoin and cryptocurrency factor exhibits a positive loading with a magnitude consistently between 0.40 and 0.60, confirming both the stablecoin and cryptocurrency markets move together and are not dominated by individual coins. However, Table~\ref{tab:pca_horn_results} highlights differences in explained variance between the two markets. On average, it seems considerably more volatility variance was captured in the cryptocurrency markets than in the stablecoin markets, suggesting that while stablecoins share systemic risks, they retain a higher degree of idiosyncratic behaviour.


\begin{table*}[htbp]
    \centering
    \caption{Horn's Parallel Analysis Results. Significant Eigenvalues highlighted in bold.}
    \resizebox{\textwidth}{!}{%
        \begin{tabular}{lccccccccc}
            \toprule
             & \multicolumn{2}{c}{PC1} & \multicolumn{2}{c}{PC2} & \multicolumn{2}{c}{PC3} & \multicolumn{2}{c}{PC4} & \\
            \cmidrule(lr){2-3} \cmidrule(lr){4-5} \cmidrule(lr){6-7} \cmidrule(lr){8-9}
            Factor & $\lambda_{corr}$ & $\lambda_{crit}$ & $\lambda_{corr}$ & $\lambda_{crit}$ & $\lambda_{corr}$ & $\lambda_{crit}$ & $\lambda_{corr}$ & $\lambda_{crit}$ & \textbf{Var. Expl. (\%)} \\
            \midrule
            $v_{\mathit{S}}$      & \textbf{2.31} & 1.07 & 0.40 & 1.02 & 0.30 & 0.99 & - & -    & \textbf{76.8\%} \\
            $\sigma_{\mathit{S, up}}$      & \textbf{1.54} & 1.08 & 0.77 & 1.02 & 0.69 & 0.99 & - & -    & \textbf{51.3\%} \\
            $\sigma_{\mathit{S, down}}$    & \textbf{1.55} & 1.08 & 0.78 & 1.02 & 0.67 & 0.98 & - & -    & \textbf{51.6\%} \\
            $v_{\mathit{C}}$      & \textbf{2.68} & 1.09 & 0.67 & 1.04 & 0.42 & 1.00 & 0.24 & 0.97 & \textbf{66.9\%} \\
            $\sigma_{\mathit{C, up}}$          & \textbf{2.55} & 1.10 & 0.69 & 1.04 & 0.47 & 1.00 & 0.28 & 0.97 & \textbf{63.8\%} \\
            $\sigma_{\mathit{C, down}}$        & \textbf{3.25} & 1.10 & 0.34 & 1.04 & 0.27 & 1.00 & 0.14 & 0.97 & \textbf{81.2\%} \\
            \bottomrule
        \end{tabular}%
    }
    \label{tab:pca_horn_results}
\end{table*}

\section{Methodology}
\label{sec:methodology}

\subsection{Causality Testing}

We start our analysis with an in-sample CGC test spanning the first four years of the dataset. Unlike traditional Granger Causality (GC) that is limited to capturing linear causality in the conditional mean, CGC detects non-linear causalities across the entire joint distribution. We examine daily, weekly, and monthly horizons to see how causality manifests over different periods, testing both directions to see how stablecoins and cryptocurrencies influence one another.

We use the same methodology outlined in the original CGC paper~\cite{copulaGranger}, made up of the following steps:

\subsubsection{\textbf{Kernel Density Estimation}} A Gaussian Kernel is used to estimate the conditional cumulative distribution function (CDF) for every factor at each time step.

\subsubsection{\textbf{Conditional Probability Transformation}} The data is transformed into uniform residuals by performing the PIT on our estimated conditional CDFs.

\subsubsection{\textbf{Bernstein Copula Approximation}} Bernstein polynomials are used for a non-parametric approximation of the joint density distribution between the factors. 

\subsubsection{\textbf{Test Statistic}} The final CGC statistic is calculated as the expected log-likelihood of the estimated density. 

\subsubsection{\textbf{Bootstrap Validation}} To assess statistical significance, we use bootstrapping to generate 200 synthetic datasets, calculating the CGC statistic for each. We reject the null hypothesis of non-causality only if the observed statistic exceeds the 95th percentile of the synthetic statistics.

\subsection{Statistical Backtesting}

To determine if stablecoin features improve predictive accuracy, we conduct OOS backtests comparing the Mean Squared Error (MSE) of a Benchmark model (cryptocurrency only) against a Challenger model (cryptocurrency and stablecoin). Models are trained on the same 4 years used in our in-sample causality test and tested on the final year of our data. We use a pairwise approach to isolate relationships between factors.

However, simple GARCH and machine learning models struggle to simultaneously handle Heteroskedasticity, asymmetric and heavy-tailed distributions, and non-linearity. To address this, we employ our GARCH-Copula-XGBoost framework:

\subsubsection{\textbf{GARCH Filtering}}

We first filter the PCA factors using the E-GARCH model outlined in Section \ref{subsec:garch}. We predict the next day at each time step and then calculate the standardised residual using the error and volatility of the prediction.

\subsubsection{\textbf{Probability Integral Transform}}

Next, we map these standardised residuals to the unit interval $(0, 1)$ using the CDF of the estimated \textit{Skew-t} distribution. This transformation isolates the dependence structure from the marginal behaviours of the individual assets, as explained in Section {\ref{subsec:copula}}.

\subsubsection{\textbf{XGBoost Copula}}

The transformed uniform residuals are fed into an XGBoost model, which acts as an estimator of the copula function.

\subsubsection{\textbf{Inverse Transformation}}

To generate the final forecast, the predicted uniform values are mapped back to the shock scale using the inverse PIT. 

During our backtest, we employ an expanding window forecasting scheme. For every day in the OOS period, PCA factors are re-calculated and GARCH parameters are re-estimated using all available history up to that point. Prior to the OOS phase, optimal XGBoost hyperparameters are selected via a grid search on the initial training set, employing a cross-validation scheme. The selected hyperparameters are fixed throughout the OOS backtest. Finally, we assess the statistical significance of any reduction in MSE using the Diebold-Mariano test outlined in Section \ref{subsec:Diebold}.

To capture the effects of stablecoin factors on our cryptocurrency factors across different time horizons, we use a variety of features in our XGBoost model. The two main feature types are the residuals and conditional volatilities derived from our E-GARCH model. For both the stablecoin and cryptocurrency factors, we use 1, 7, and 30-day lags as well as 7 and 30 day moving averages of these factors to capture all horizons.

\subsection{Volatility Targeting}
\label{subsec:VolTargetMethod}

While the statistical backtest establishes the effect of stablecoin factors on the predictive error of cryptocurrency factors, we must determine if this accuracy translates into real economic value. We are still using Principal Components to predict other principal components, so now we must convert these predictions into actionable decisions for individual assets. To bridge this gap, we construct a dynamic portfolio of cryptocurrency assets where the allocation for each asset is derived from the principal component.

We employ the same GARCH-Copula-XGBoost framework as defined in the statistical test. However, we extend the methodology to support a volatility targeting strategy~\cite{volatilityTargeting} to manage the risk of a portfolio made up of our selected cryptocurrencies. The backtest runs on the same year as before, with the same expanding window and hyperparameter tuning approach. The core of our strategy relies on forecasting the asymmetry between upside and downside market volatility. We summarise the strategy below:

\subsubsection{\textbf{Signal Generation}}

We define the Net Volatility Signal $S_t$ as the normalised difference between the forecasted upside and downside volatilities. 

For each day $t$, we calculate the upside and downside volatilities for all assets and calculate the average for each ($\bar{\sigma}_{\mathit{dir,t}}, \mathit{dir} \in \{\mathit{up,down}\}$) using the raw price data. The XGBoost model then predicts the shock $\Delta \hat{y}$ for the principal component of each volatility and the final volatility forecasts are reconstructed as:
\begin{equation}
\hat{\sigma}_{\mathit{dir}, t+1} = \bar{\sigma}_{\mathit{dir}, t} + \Delta \hat{y}_{\mathit{dir}, t+1} \\
\end{equation}
\begin{equation}
\hat{\sigma}_{\mathit{daily}, t+1} = \hat{\sigma}_{\mathit{up}, t+1} + \hat{\sigma}_{\mathit{down}, t+1}
\end{equation}
The signal is then derived as the ratio of the volatility spread to the total risk:
\begin{equation}
S_{t+1} = \frac{\hat{\sigma}_{\mathit{up}, t+1} - \hat{\sigma}_{\mathit{down}, t+1}}{\hat{\sigma}_{\mathit{up}, t+1} + \hat{\sigma}_{\mathit{down}, t+1}}
\end{equation}

A positive signal suggests that upside variance is the dominant driver of the market factor, prompting a buy position.

\subsubsection{\textbf{Portfolio Construction and Weighting}}

We use the principal component loadings from Table~\ref{tab:pca_loadings_combined} to weight our portfolio. More specifically, we give a higher allocation to assets more sensitive to upside volatility than downside volatility. We compute a weighting ratio for each asset $i$:
\begin{equation}
R_i = \frac{Loading_{i, \mathit{up}}}{Loading_{i, \mathit{down}}}
\end{equation}

Then determine the final portfolio weights $w_i$ by normalising these ratios:
\begin{equation}
w_i = \frac{R_i}{\sum_{j=1}^{N} R_j}
\end{equation}

This ensures that the portfolio tilts towards assets that contribute more to the upside volatility rather than the downside.

\subsubsection{\textbf{Position Sizing and Risk Management}}

We implement a volatility targeting framework to manage risk. For a given annual volatility target ($\sigma_{target}$), the base exposure level is calculated relative to the forecasted annualised risk:
\begin{equation}
\text{Exp}_{\mathit{base}} = \frac{\sigma_{\mathit{target}}}{\hat{\sigma}_{\mathit{ann}, t+1}}
\end{equation}
\begin{equation}
\hat{\sigma}_{\mathit{ann}, t+1} = \hat{\sigma}_{\mathit{daily}, t+1} \times \sqrt{366}
\end{equation}

where $366$ is the number of days in 2024. To further refine the strategy, we apply a confidence multiplier based on the historical strength of our signal. We calculate the z-score ($z_S$) of the current signal $S_{t+1}$ relative to its rolling 60-day history. The final gross exposure is adjusted using a hyperbolic tangent function to smooth extreme signals:
\begin{equation}
\text{Multiplier}_t = 1 + \tanh(z_S)
\end{equation}
\begin{equation}
\text{Exp}_{\mathit{total}} = \text{Exp}_{\mathit{base}} \times \text{Multiplier}_t
\end{equation}
This enables us to leverage when the model's signal is statistically strong and decreases it when the signal is weak.

\subsubsection{\textbf{Performance Evaluation}}

As established in the statistical analysis, we retain the Benchmark and Challenger model configurations. However, to assess the economic value of our framework, we introduce a Naive baseline strategy. Whilst the Challenger model may outperform the Benchmark, both models could be poor overall. The Naive strategy acts as a baseline to determine if our models yield real economic value in comparison. It employs an equal-weighted allocation across the assets and sizes positions using a trailing volatility average.

We use the two challenger models for cryptocurrency upside and downside volatility to predict a full picture of volatility. We then use these predictions to calculate our signal. We run two volatility targeting tests at the $20\%$ and $50\%$ levels. We choose the former due to this being the industry standard for volatility targeting on cryptocurrency indices~\cite{cryptoVolIndices}. We choose the latter to align with the volatility if we were to Buy\&Hold in our test period so we can compare our model's performance against the passive strategy. Transaction costs are modelled at 1 basis point ($0.01\%$) per turnover (difference in portfolio weights between today and yesterday) to account for daily portfolio rebalancing. Performance across all strategies is evaluated using Annualised Return, Sortino Ratio, and Maximum Drawdown. We do not consider market impact.

\section{Results}
\label{sec:results}

\subsection{Causality Testing}

Table \ref{tab:granger_causality} shows the results of our CGC tests using stablecoin factors as the causer and cryptocurrency factors as the target. The values are the p-values for each test, where \textbf{$<$0.005} means that the observed CGC statistic was to the right of our entire synthetic distribution. Fig. \ref{fig:copula_null} shows an example synthetic CGC distribution to highlight what this looks like.

\begin{figure}[htbp]
    \centering
    \includegraphics[width=\columnwidth]{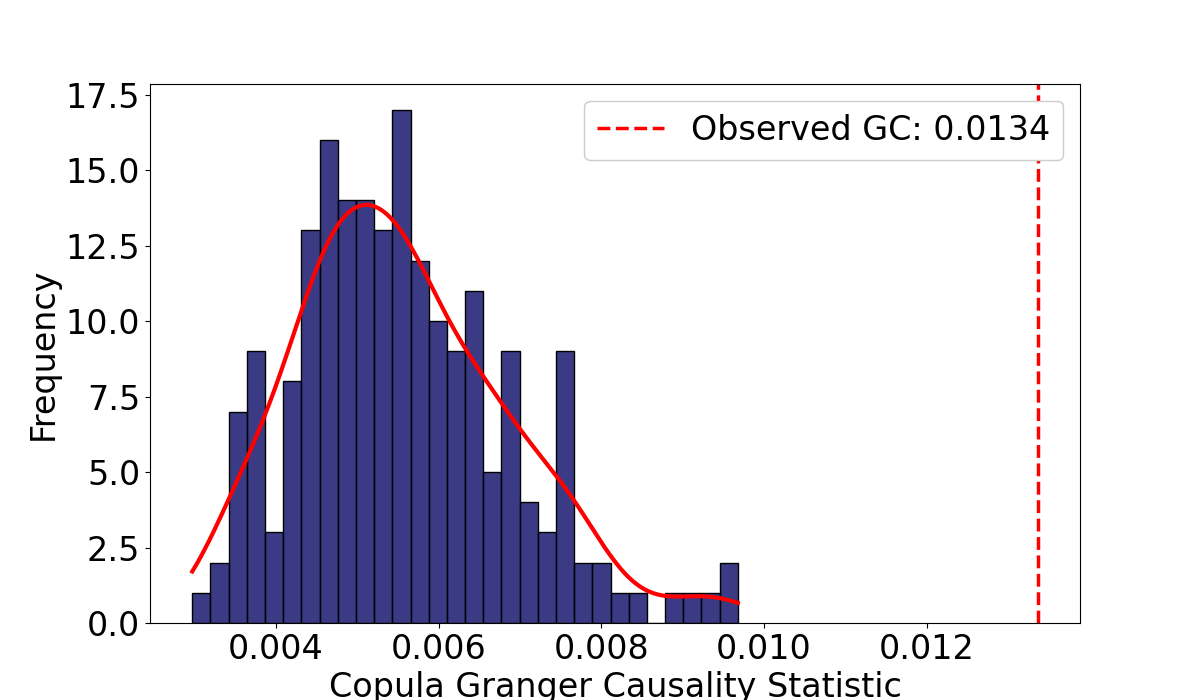} 
    \caption{Bootstrapped CGC Distribution for Stablecoin Upside Volatility causing Cryptocurrency Upside Volatility on the Weekly Horizon.}
    \label{fig:copula_null}
\end{figure}

\begin{table}[htbp]

\centering

\caption{Copula Granger Causality Test Results for Both Directions. Significant results highlighted in bold.}

\label{tab:granger_causality}

\renewcommand{\arraystretch}{1}

\resizebox{\columnwidth}{!}{%

    \begin{tabular}{lrrr}

    \toprule

    Stablecoin $\rightarrow$ Crypto & Daily & Weekly & Monthly \\

    \midrule

    $\sigma_{\mathit{down}}$ $\rightarrow$ $\sigma_{\mathit{down}}$ & \textbf{$<$0.005} & \textbf{$<$0.005} & \textbf{0.005} \\

    $\sigma_{\mathit{up}}$ $\rightarrow$ $\sigma_{\mathit{down}}$   & \textbf{0.05}     & \textbf{$<$0.005} & \textbf{$<$0.005} \\

    $v$ $\rightarrow$ $\sigma_{\mathit{down}}$                      & 0.155             & 0.065             & \textbf{$<$0.005} \\

    $\sigma_{\mathit{down}}$ $\rightarrow$ $\sigma_{\mathit{up}}$   & \textbf{$<$0.005} & \textbf{$<$0.005} & \textbf{0.01} \\

    $\sigma_{\mathit{up}}$ $\rightarrow$ $\sigma_{\mathit{up}}$     & \textbf{$<$0.005} & \textbf{$<$0.005} & 0.200 \\

    $v$ $\rightarrow$ $\sigma_{\mathit{up}}$                        & 0.740             & \textbf{0.02}     & \textbf{$<$0.005} \\

    \midrule

    Crypto $\rightarrow$ Stablecoin & Daily & Weekly & Monthly \\

    \midrule

    $\sigma_{\mathit{down}}$ $\rightarrow$ $\sigma_{\mathit{down}}$ & \textbf{$<$0.005} & \textbf{$<$0.005} & \textbf{0.01} \\

    $\sigma_{\mathit{up}}$ $\rightarrow$ $\sigma_{\mathit{down}}$   & \textbf{$<$0.005} & \textbf{$<$0.005} & \textbf{$<$0.005} \\

    $v$ $\rightarrow$ $\sigma_{\mathit{down}}$                      & 0.300             & 0.695             & 0.080 \\

    $\sigma_{\mathit{down}}$ $\rightarrow$ $\sigma_{\mathit{up}}$   & \textbf{$<$0.005} & \textbf{0.005}    & \textbf{$<$0.005} \\

    $\sigma_{\mathit{up}}$ $\rightarrow$ $\sigma_{\mathit{up}}$     & \textbf{$<$0.005} & \textbf{$<$0.005} & \textbf{$<$0.005} \\

    $v$ $\rightarrow$ $\sigma_{\mathit{up}}$                        & 0.240             & 0.055             & 0.310 \\

    \bottomrule

    \end{tabular}

}

\end{table}

The results show that stablecoin factors significantly cause for cryptocurrency factors across almost all factor combinations and time horizons. Stablecoin downside volatility emerges as the most influential factor, exhibiting statistically significant causality across all factors and horizons. Similarly, upside volatility shows causality in almost all factors and areas but appearing to diminish on the monthly horizon for cryptocurrency upside volatility. In contrast, stablecoin volume acts as a long-term signal, lacking significance on a daily horizon but becoming significant on weekly and monthly horizons. We see very similar results for the reverse direction, with cryptocurrency both volatilities causing both stablecoin volatilities across all horizons. Interestingly, cryptocurrency volume has no significant causality for any stablecoin factor.  

\subsection{Statistical Backtest}

Table \ref{tab:MSE} shows the results of our pairwise OOS MSE backtests. A positive MSE reduction means the challenger model reduced the MSE compared to the benchmark model. We also include its performance against a simple E-GARCH model to show our framework's improved performance.

\begin{table}[htbp]
\centering
\caption{MSE Backtest Results for the Challenger Model against the Benchmark and E-GARCH models. Significant Results Highlighted in Bold.}
\label{tab:MSE}
\resizebox{\columnwidth}{!}{
    \begin{tabular}{lrrrr}
    \toprule
    & \multicolumn{2}{c}{vs. Benchmark} & \multicolumn{2}{c}{vs. EGARCH} \\
    \cmidrule(lr){2-3} \cmidrule(lr){4-5}
    Stablecoin $\rightarrow$ Crypto & \begin{tabular}[c]{@{}r@{}}MSE Red.\\ (\%)\end{tabular} & DM $p$-val & \begin{tabular}[c]{@{}r@{}}MSE Red.\\ (\%)\end{tabular} & DM $p$-val \\
    \midrule
    $\sigma_{\mathit{down}}$ $\rightarrow$ $\sigma_{\mathit{down}}$ & \textbf{3.44}  & \textbf{0.010}     & \textbf{25.74} & \textbf{0.002} \\
    $\sigma_{\mathit{up}}$ $\rightarrow$ $\sigma_{\mathit{down}}$   & 0.87           & 0.086              & \textbf{23.64} & \textbf{$<$0.001} \\
    $v$ $\rightarrow$ $\sigma_{\mathit{down}}$                      & \textbf{4.62}  & \textbf{$<$0.0001} & \textbf{26.31} & \textbf{$<$0.001} \\
    $\sigma_{\mathit{down}}$ $\rightarrow$ $\sigma_{\mathit{up}}$   & -1.92          & 0.109              & 0.22           & 0.923 \\
    $\sigma_{\mathit{up}}$ $\rightarrow$ $\sigma_{\mathit{up}}$     & \textbf{9.48}  & \textbf{$<$0.0001} & \textbf{11.38} & \textbf{$<$0.001} \\
    $v$ $\rightarrow$ $\sigma_{\mathit{up}}$                        & \textbf{2.47}  & \textbf{0.002}     & 4.50           & 0.056 \\
    \bottomrule
    \end{tabular}
}
\end{table}

The backtest reveals a divergence between in-sample causality and OOS predictive power, particularly in downside volatility. While the causality tests identified stablecoin downside volatility as a dominant causal driver across all factors and horizons, these signals failed to generalise out-of-sample. More specifically, stablecoin downside volatility marginally reduced MSE for cryptocurrency downside volatility but deteriorated performance for upside volatility. Conversely, stablecoin upside volatility demonstrated significant MSE reduction, seeing a $9.48\%$ reduction in MSE for cryptocurrency upside volatility. Moreover, stablecoin volume delivered significant MSE reductions for both Upside and Downside volatility.

Fig. \ref{fig:feature_importance} shows the Gain for the top 10 features in the Challenger model for cryptocurrency upside volatility, which uses both stablecoin and cryptocurrency features. We can clearly see stablecoin features dominate, with the top four features all derived from stablecoin data. Interestingly, we saw no significant causality on the monthly horizon yet we see the 30-day moving average as the second most important feature in our model. This suggests that the information gain does not come directly from the monthly horizon itself but rather from its interaction with other factors.

\begin{figure}[htbp]
    \centering
    \includegraphics[width=\columnwidth]{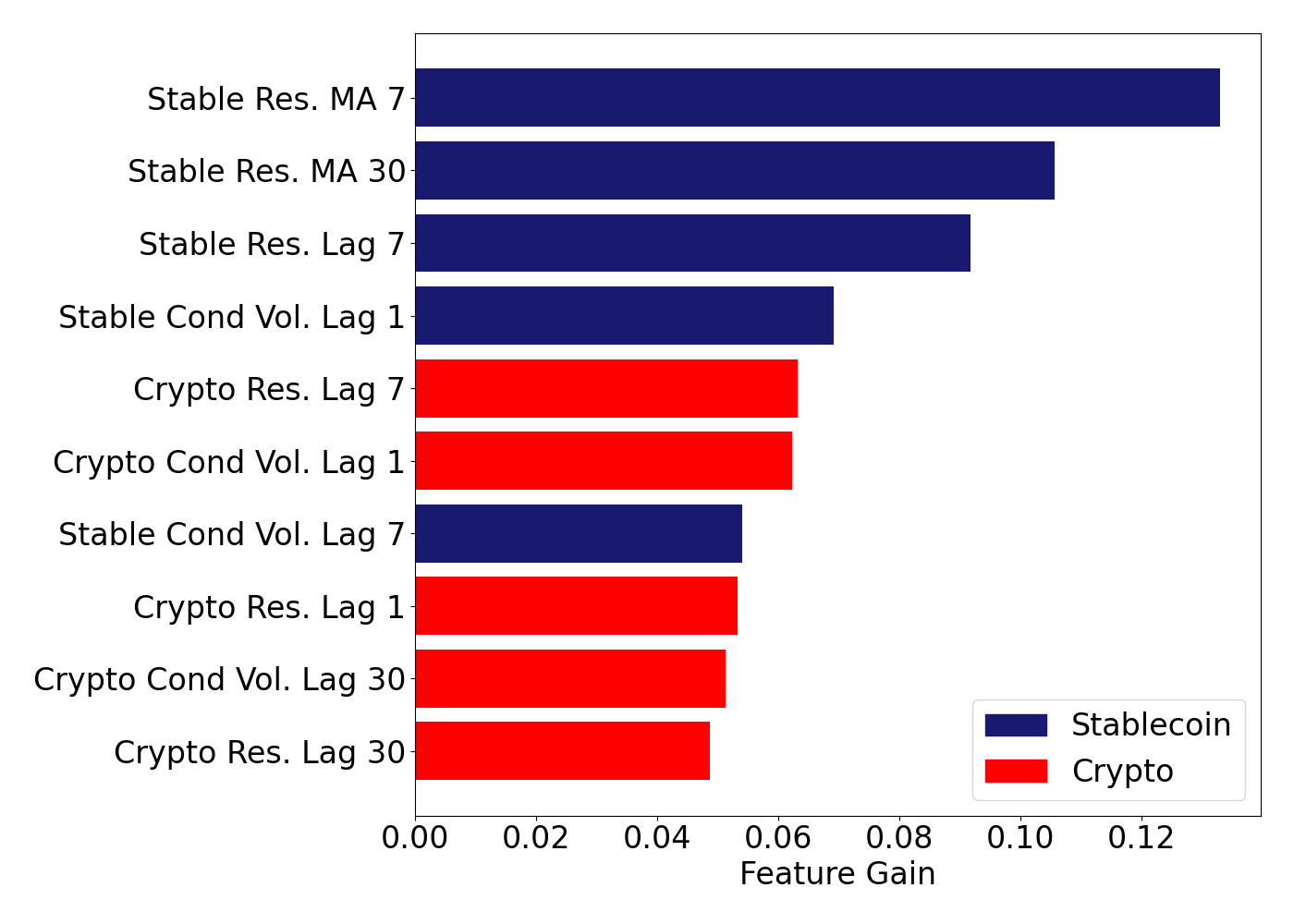} 
    \caption{Top 10 Feature Gains for Stablecoin Upside Volatility Predicting Crypto Upside Volatility.}
    \label{fig:feature_importance}
\end{figure}

\subsection{Volatility Targeting}

Table \ref{tab:vol_targeting_results} contains the results of our backtests. Our results demonstrate that using stablecoin-driven signals significantly improves performance for both volatility targets. At the $20\%$ target, the Challenger model achieves a Sortino ratio over 40\% larger than the benchmark. with a $6\%$ increase in returns and a $3\%$ decrease in Max Drawdown. Moreover, at the $50\%$ level, the Challenger model achieves a Sortino ratio over 30\% larger than the Buy\&Hold strategy, with an $11\%$ increase in returns and $9\%$ decrease in max drawdown.

\begin{table}[htbp]
    \centering
    \caption{Volatility Targeting Performance Metrics.}
    \label{tab:vol_targeting_results}
    \resizebox{\columnwidth}{!}{%
        \begin{tabular}{lccccc}
            \toprule
            Strategy & $\sigma_{\mathit{target}}$ & Ann. Ret. & Ann. Vol. & \makecell{Max \\ Drawdown} & \makecell{Sortino \\ Ratio} \\
            \midrule
            Buy \& Hold & - & 89.0\% & 51.4\% & -33.0\% & 2.47 \\
            \midrule
            Benchmark & 20\% & 40.8\% & 24.3\% & -15.8\% & 1.96 \\
            \textbf{Challenger} & \textbf{20\%} & \textbf{46.6\%} & \textbf{22.3\%} & \textbf{-12.9\%} & \textbf{2.77} \\
            Naive & 20\% & 16.6\% & 19.7\% & -18.6\% & 1.10 \\
            \midrule
            
            Benchmark & 50\% & 95.8\% & 46.1\% & -26.9\% & 3.04 \\
            \textbf{Challenger} & \textbf{50\%} & \textbf{100.4\%} & \textbf{44.7\%} & \textbf{-24.1\%} & \textbf{3.38} \\
            Naive & 50\% & 54.0\% & 38.9\% & -30.2\% & 2.00 \\
            \bottomrule
        \end{tabular}%
    }
\end{table}

For further context, Fig. \ref{fig:PnL} shows model performance through the test period. Generally speaking, our test period was bullish, with bearish and flat periods in the middle of the year. Our model maintained an upward trend during this bearish period, suggesting our model is effective all market conditions.

\begin{figure}[htbp]
    \centering
    \includegraphics[width=\columnwidth]{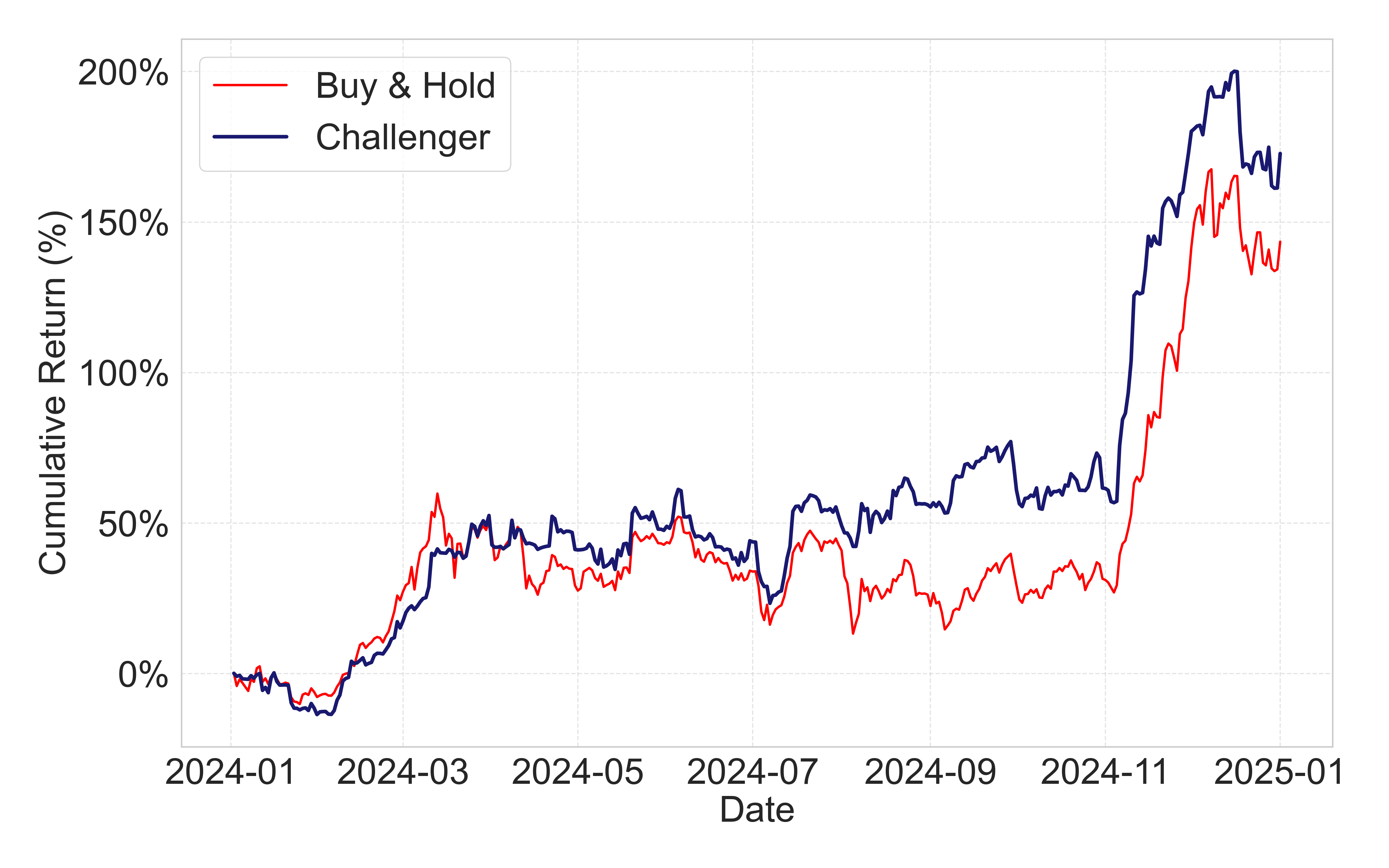} 
    \caption{Model Performance Over Time}
    \label{fig:PnL}
\end{figure}
\section{Discussion}
\label{sec:discussion}

Upside volatility in stablecoins indicates movement of liquidity into the stablecoin layer, attributed to using stablecoins as either a safe-haven in poor market conditions or as a dry powder waiting to be used in an impending rally. Our research captures the dry powder effect, where stablecoin upside volatility leads upside volatility in cryptocurrencies. With this in mind, however, we would also expect to see stablecoin downside volatility lead cryptocurrency upside volatility, as the dry powder must eventually be liquidated to instigate the rally. Instead, we see a deteriorated performance. This divergence is likely due to our daily timescale - downside volatility would precede cryptocurrency upside volatility almost instantaneously when investors start buying cryptocurrency so would be captured on an intra-day timeframe. This is a limitation of this study as we are confined to using daily data.

In contrast, while upside volatility acts as a directional signal, stablecoin volume saw significant results across the whole volatility picture, indicating the stablecoin market's role as the engine for the broader cryptocurrency market and a measure of the dry powder available. Moreover, its significance on the weekly and monthly horizons suggests the accumulation of dry powder is a gradual process. Rather than driving instantaneous shocks, stablecoin volume seems to dictate the markets overall capacity for volatility on longer horizons.

The regulatory landscape seems to reinforce the dry powder dynamic. The GENIUS Act pioneered the "payment stablecoins" model, with similar frameworks emerging in other countries such as the United Kingdom~\cite{UKStablecoinsAct} and Canada~\cite{CanadaStablecoinAct}.
The act prohibits yield generation and mandates 1:1 reserves backing without the option to rehypothecate~\cite{GENIUS2025}. The prohibition of yield generation enforces the role of stablecoins as a medium of exchange only, maintaining its role as dry powder. Moreover, the ban on rehypothecation ensures stablecoins remain highly liquid and accessible. Again, this enforces the role of stablecoins as the engine for the cryptocurrency market, ensuring liquidity is present to instigate rallies.

Whilst prohibited, yield-generating stablecoins are exploding in popularity, with market capitalisation up $300\%$ Year-over-Year following the GENIUS Act~\cite{yieldStablecoinMarketCap} . Common types include delta-neutral currencies like USDe~\cite{usdeMarketCap}, Real World Asset (RWA)-backed like USDM for US treasury bills~\cite{usdmMarketCap}, and tokenised treasuries like BUIDL for US government securities~\cite{blackrockBUIDL}. As the purpose of these stablecoins goes beyond a medium of exchange and introduce exogenous factors, it is likely that these types of stablecoins will not act as dry powder.

\section{Conclusion}
\label{sec:conclusion}

This study presents a quantitative analysis of the transmission of volatility and activity from stablecoins to cryptocurrency markets. We used copula-based approaches for in-sample and out-of-sample testing, obtaining evidence that stablecoins act as leading indicators for broader cryptocurrency market volatility. Using our novel GARCH-Copula-XGBoost framework, we provide evidence for the stablecoin market's role as a dry powder, showing significant MSE reductions for cryptocurrency volatility when employing stablecoin upside volatility and volume in our models. Finally, we show how to translate these findings into economic value with our volatility targeting model to manage risk. This paper highlights the role that stablecoins have, going beyond their a passive pegging mechanism to acting as the engine for the ecosystem. 

Future work includes testing the reverse direction to see how cryptocurrencies lead stablecoins, with the potential to capture the safe-haven dynamic where cryptocurrency downside volatility leads stablecoin upside volatility. Moreover, the research could be repeated on intra-day time frames, aiming to uncover more instantaneous relationships and the missing link between stablecoin downside volatility and cryptocurrency upside volatility we previously discussed.


%
%
\bibliographystyle{splncs04}
\bibliography{bibliography}
%

\end{document}